\documentclass[fleqn,usenatbib,usedcolumn]{mnras}
\usepackage[british]{babel}
\usepackage{newtxtext}
\usepackage[slantedGreek]{newtxmath}
\usepackage[T1]{fontenc}
\usepackage{graphicx, mathrsfs}

\usepackage{hyperref}
\hypersetup{
    pdfnewwindow=true,      
    colorlinks=true,       
    linkcolor=blue,          
    citecolor=cyan,        
    filecolor=blue,      
    urlcolor=blue           
}

\def\prd{PRD}
\def\apj{ApJ}                 
\def\apjl{ApJL}               
\def\apjs{ApJS}               
\def\mnras{MNRAS}             
\def\aap{A\&A}                
\def\aj{AJ}                   
\def\nat{Nature}              

\title[Eccentric Black Hole Mergers in LISA]{Black Hole Mergers From Globular Clusters Observable by LISA I: Eccentric Sources Originating From Relativistic $N$-body Dynamics}

\author[Samsing, D'Orazio]{
Johan Samsing$^{1,}$\thanks{jsamsing@princeton.edu; daniel.dorazio@cfa.harvard.edu},
Daniel J. D'Orazio$^{2}$
\\
$^{1}$Department of Astrophysical Sciences, Princeton University, Peyton Hall, 4 Ivy Lane, Princeton, NJ 08544, USA\\
$^{2}$Department of Astronomy, Harvard University, 60 Garden Street Cambridge, MA 01238, USA}

\date{Accepted XXX. Received YYY; in original form ZZZ}

\pubyear{2018}

\begin{document}
\label{firstpage}
\pagerange{\pageref{firstpage}--\pageref{lastpage}}
\maketitle

\begin{abstract}

We show that nearly half of all binary black hole (BBH) mergers dynamically
assembled in globular clusters have measurable eccentricities ($e>0.01$) in
the LISA band ($10^{-2}$ Hz), when General Relativistic corrections are
properly included in the $N$-body evolution. If only Newtonian gravity is
included, the derived fraction of eccentric LISA sources is significantly
lower, which explains why recent studies all have greatly underestimated this
fraction. Our findings have major implications for how to observationally
distinguish between BBH formation channels using eccentricity with LISA, which
is one of the key science goals of the mission. We illustrate that the
relatively large population of eccentric LISA sources reported here originates
from BBHs that merge between hardening binary-single interactions inside their
globular cluster. These results indicate a bright future for using LISA to
probe the origin of BBH mergers.

\end{abstract}



\section{Introduction}

Gravitational waves (GWs) from merging binary black holes (BBHs) have been
observed \citep{2016PhRvL.116f1102A, 2016PhRvL.116x1103A, 2016PhRvX...6d1015A,
2017PhRvL.118v1101A, 2017PhRvL.119n1101A}; however with the sparse sample
collected to far, it is not clear where and how these BBHs formed in our
Universe. From a theoretical perspective, several formation channels have been
suggested including isolated field binaries \citep{2012ApJ...759...52D,
2013ApJ...779...72D, 2015ApJ...806..263D, 2016ApJ...819..108B,
2016Natur.534..512B}, dense stellar clusters \citep{2000ApJ...528L..17P,
2010MNRAS.402..371B, 2013MNRAS.435.1358T, 2014MNRAS.440.2714B,
2015PhRvL.115e1101R, 2016PhRvD..93h4029R, 2016ApJ...824L...8R,
2016ApJ...824L...8R, 2017MNRAS.464L..36A, 2017MNRAS.469.4665P}, single-single
GW captures of primordial BHs \citep{2016PhRvL.116t1301B, 2016PhRvD..94h4013C,
2016PhRvL.117f1101S, 2016PhRvD..94h3504C}, active galactic nuclei discs
\citep{2017ApJ...835..165B,  2017MNRAS.464..946S, 2017arXiv170207818M},
galactic nuclei \citep{2009MNRAS.395.2127O, 2015MNRAS.448..754H,
2016ApJ...828...77V, 2016ApJ...831..187A, 2017arXiv170609896H}, and very
massive stellar mergers \citep{Loeb:2016, Woosley:2016, Janiuk+2017,
DOrazioLoeb:2017}. Although these proposed pathways seem to give rise to
similar merger rates and observables, recent work interestingly suggests that
careful measurements of the BBH orbital eccentricity and relative spins might
be the key to disentangling them. For example, BBH mergers forming as a result
of field binary evolution are likely to have correlated spin orientations,
except if a third object is bound and the three objects form a hierarchical
triple \citep[\textit{e.g.},][]{2017ApJ...846L..11L, 2017arXiv171107142A}, whereas BBH
mergers forming in clusters are expected to have randomized orientations due
to frequent exchanges \citep[\textit{e.g.},][]{2016ApJ...832L...2R}. Regarding
eccentricity, it was recently shown by \cite{2017arXiv171107452S} that
$\approx 5\%$ of all BBH mergers forming in globular clusters (GCs) are likely
to have a notable eccentricity ($ > 0.1$) when entering the observable range
of the `Laser Interferometer Gravitational-Wave Observatory' (LIGO).  As
argued by \cite{2017arXiv171107452S}, this population originates from GW
capture mergers forming in chaotic three-body interactions
\citep[\textit{e.g.},][]{2006ApJ...640..156G, 2014ApJ...784...71S} during classical
hardening, which explains why all recent Newtonian $N$-body studies have
failed in resolving the correct fraction. In fact, it was analytically derived
in \cite{2017arXiv171107452S} that a Newtonian $N$-body code will always
result in a rate of eccentric mergers that is $\approx 100$ times lower
compared to the (correct) General Relativistic (GR) prediction. These results
were recently confirmed by \cite{2017arXiv171204937R} and
\cite{2018ApJ...855..124S}, using data from simulated GCs. As isolated BBH
mergers forming in the field are expected to be circular when entering the
LIGO band,
these studies show that eccentricity could play a key role in
distinguishing formation channels from each other.

In this paper we study how BBH mergers that form dynamically in GCs distribute
and evolve in the sensitivity band of the proposed `Laser Interferometer Space
Antenna' mission \citep[LISA;][]{2017arXiv170200786A}, when GR effects are
included in the dynamical modeling. Dissipative effects, such as GW emission,
which usually are modeled using the post-Newtonian (PN) formalism
\citep[\textit{e.g.},][]{2014LRR....17....2B}, have previously been shown to
play a crucial role in resolving eccentric LIGO populations
\citep[\textit{e.g.},][]{2018ApJ...853..140S}; however, the possible effects related to
LISA have not yet been properly studied. Our motivation is to explore what can
be learned about where and how BBH mergers form in our Universe from a LISA
mission; we identify possible observable differences between different BBH
populations formed in GCs compared to those formed in the field. Motivated
by previous studies, we focus in this paper on the eccentricity distribution.
We note that the recent work by
\cite{2016ApJ...830L..18B} did indeed look into this; however, the data used
for that study did not include GR effects, which we in this paper show are
extremely important.

Using a semi-analytical approach, we find that $\approx 4$ times more BBH
mergers will appear eccentric ($>0.01$) in the LISA band ($10^{-2}$ Hz)
compared to the results reported by \cite{2016ApJ...830L..18B}, when GR
effects are included. This leads to the exciting conclusion that about $40\%$
of all GC BBH mergers are expected to have a measurable eccentricity in the
LISA band, whereas a field BBH population in comparison will have $\approx
0\%$. As we describe, the merger population that leads to this increase in
eccentric LISA sources, originates from BBHs that merge \emph{between} their
hardening binary-single interactions inside their GC
\citep[\textit{e.g.},][]{2017arXiv171204937R}. This population was not included in the
recent study by \cite{2018arXiv180208654S}, which focused solely on the BBH
mergers forming \emph{during} the binary-single interactions. These BBH
mergers where shown to elude the LISA band, and joint observations with LIGO
are therefore necessary to tell their GC origin. The fact that BBH mergers can
be jointly observed by LISA and LIGO was recently pointed out by
\cite{2016PhRvL.116w1102S} and \cite{Seto:2016}. Discussions on BBH merger
channels and eccentricity distributions relevant for LISA were presented in
\cite{2016PhRvD..94f4020N} and \citep{2017MNRAS.465.4375N}. However, we note
again that all previous studies have greatly underestimated the fraction of
eccentric LISA sources from GCs, mainly due to the omitted GR effects in the
data set derived in \cite{2016PhRvD..93h4029R}. It would be interesting to see
how the results presented in this paper affect those previous studies.

The paper is organized as follows. In Section \ref{sec:BBH mergers in the LISA
band} we describe the approach we use for modeling the dynamical evolution of
BBHs inside GCs and their path towards merger, when GR effects are included in
the problem. Our main results are discussed in Section \ref{sec:Results},
where we show for the first time that, with the inclusion of GR effects, nearly
half of all BBH mergers forming in GCs are expected be eccentric in LISA. We
conclude our study in Section \ref{sec:Conclusions}.

\section{Black Hole Dynamics in Globular Clusters}\label{sec:BBH mergers in the LISA band}

\begin{figure}
\centering
\includegraphics[width=\columnwidth]{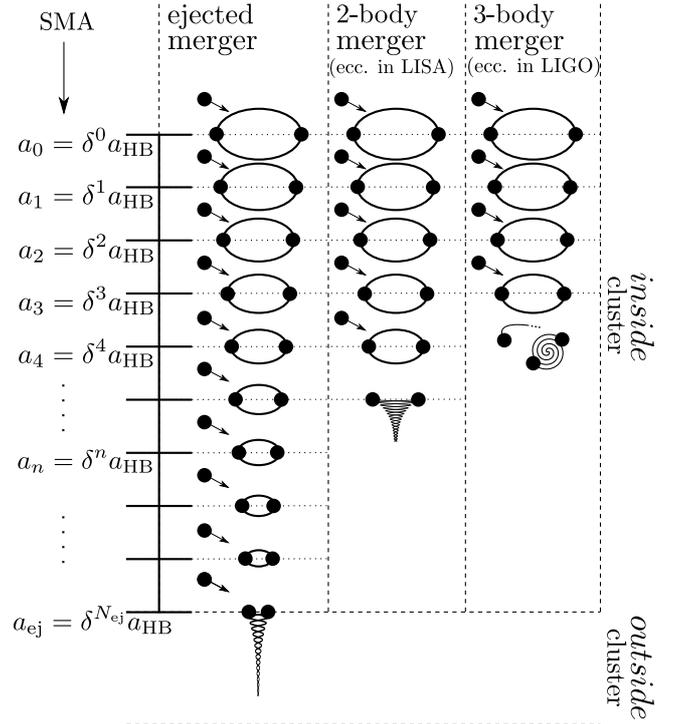}
\caption{
The graphics in the three columns above illustrate the three different dynamical pathways for merging BBHs to form, each of which result in a different type of GW merger.
The horizontal steps from top to bottom illustrate the stepwise decrease in the BBH's SMA due to hardening binary-single interactions, which progresses as
${\delta}^{0}a_{\rm HB}$, ${\delta}^{1}a_{\rm HB}$, ..., until a merger or an ejection takes place.
The illustration complements the description of our model from Section \ref{sec:BBH mergers in the LISA band}. In short, our model
assumes the BBH in question starts with an SMA $ = a_{\rm HB}$, after which it hardens through binary-single interactions,
each of which leads to a decrease in its SMA from $a$ to ${\delta}a$. This hardening continues until the SMA
reaches $a_{\rm ej}$, below which the BBH will be ejected from the GC through three-body recoil. If the BBH merges outside the GC within a Hubble time,
we label it an `ejected merger' (left column). The ejected merger progenitors form via interactions involving Newtonian gravity alone; however, when GR effects
are included, the BBH can also merge inside the cluster, before ejection takes place \citep[e.g.][]{2017arXiv171107452S, 2017arXiv171204937R}. This can
happen either \textit{between} or \textit{during} its hardening interactions, outcomes we refer to as a `2-body merger' (middle column) and a `3-body merger' (right column), respectively.
All previous studies on the eccentricity distribution of LISA sources have only considered the `ejected mergers'; however, as we show in this paper, 
the `$2$-body mergers' clearly dominate the eccentric population observable by LISA ($e>0.01$ at $10^{-2}$ Hz).
In comparison, the `3-body mergers' dominate the eccentric population observable by LIGO ($e>0.1$ at $10$ Hz).}
\label{fig:illustration}
\end{figure}

In this section we describe the new approach we use in this paper for
estimating the distribution of GW frequency, $f_{\rm GW}$, and eccentricity,
$e$, of BBH mergers forming in globular clusters (GCs). Using this, we explore
the possible observable differences between different  BBH populations forming
in GCs and those forming in the field for an instrument similar to LISA, and
the role of GR in that modeling. As described in the Introduction, in the
recent work by \cite{2016ApJ...830L..18B} it was claimed that $\approx 10\%$
of the GC mergers will have an eccentricity $>0.01$ at $10^{-2}$ Hz, compared
to $\approx 0\%$ for the field population. However, the simulations used is
\cite{2016ApJ...830L..18B} did not treat the relativistic evolution of BBHs
inside the GC correctly, which essentially prevented BBHs to merge inside
their GCs (see \cite{2017arXiv171204937R} for a description).   To improve on
their study we combine in the sections below a simple Monte Carlo (MC) method
with the analytical framework from \cite{2017arXiv171107452S} to estimate what
the actual BBH eccentricity distribution is expected to be in the LISA band,
taking into account that BBHs can form both during and between hardening
binary-single interactions \citep[\textit{e.g.},][]{2017arXiv171204937R}. Although our approach
is highly simplified, we do clearly find that GR effects play a central role in such a study. Figure \ref{fig:illustration}
schematically illustrates our dynamical model described below.

\subsection{Binary Black Holes Interacting in Clusters}\label{sec:Model}

We assume that the dynamical history of a BBH in a GC from its formation to final merger follows the idealized model described in \cite{2017arXiv171107452S},
in which it first forms dynamically at the hard-binary (HB) limit \citep[\textit{e.g.},][]{Heggie:1975uy, 1976A&A....53..259A, Hut:1983js}, after which
it hardens through equal mass three-body interactions. Each interaction
leads to a fixed decrease in its semi-major axis (SMA) from $a$ to $\delta a$, where the average value of $\delta$ is
$7/9$ using the distributions from \cite{Heggie:1975uy}, as shown by \cite{2017arXiv171107452S}. For
simplicity we will use this value of $\delta$ for our modeling.
The BBH will harden in this way until it either merges inside the GC, or its three-body recoil velocity exceeds the escape velocity of the GC, $v_{\rm esc}$, after
which it escapes. In this model, such an `ejection' can only happen if the SMA of the BBH is
below the following characteristic value \citep{2017arXiv171107452S},
\begin{equation}
a_{\rm ej} \approx \frac{1}{6} \left(\frac{1}{\delta} - 1\right) \frac{Gm}{v_{\rm esc}^2},
\end{equation}
where $m$ is the mass of one of the three interacting (assumed equal mass) BHs.
The mergers that are normally considered, using Newtonian prescriptions, are the BBHs that will merge \emph{outside} the GC,
i.e the subset of the ejected BBHs that has a GW lifetime that is less than the Hubble time, $t_{\rm H}$.
However, when GR effects are included, a BBH can also merge \emph{inside} the GC in at least two different ways \citep[\textit{e.g.},][]{2017arXiv171107452S, 2017arXiv171204937R}:
The first way is \emph{between} its hardening binary-single interactions -- a merger type we will refer to in short as a `2-body merger' (2b).
A BBH will undergo such a merger if its GW lifetime is shorter than the time it takes for the next interaction to occur.
The second way is \emph{during} its hardening binary-single interactions -- a merger type we will refer to in short as a `3-body merger' (3b). Such a merger
occurs if two of the three interacting BHs undergo a two-body GW capture merger during the chaotic evolution of the three-body system \citep{2006ApJ...640..156G, 2014ApJ...784...71S}.

These three different types of mergers (ejected merger, 2-body merger, and 3-body merger) arise, as described, from different mechanisms that each have their own characteristic
time scale (Hubble time, binary-single encounter time, three-body orbital time), which explains why they give rise to different distributions in GW frequency and eccentricity,
as will be shown in Section \ref{sec:Results}. Below we describe how we construct these distributions from our simple model.

\subsection{Deriving Eccentricity and GW Frequency Distributions}

We start by considering two BHs each with mass $m$, in a binary with SMA equal to their HB value given by \citep[\textit{e.g.},][]{Hut:1983js},
\begin{equation}
a_{\rm HB} \approx \frac{3}{2}\frac{Gm}{v_{\rm dis}^{2}},
\end{equation}
where $v_{\rm dis}$ is the velocity dispersion of the interacting BHs.
As described in Section \ref{sec:Model} and shown in Figure \ref{fig:illustration}, we assume that the dynamical evolution of this
BBH is governed by isolated binary-single interactions that lead to a stepwise decrease in its SMA as follows,
${\delta}^{0}a_{\rm HB}, {\delta}^{1}a_{\rm HB}, {\delta}^{2}a_{\rm HB}, ..., {\delta}^{n}a_{\rm HB}, ..., $ until ${\delta}^{N_{\rm ej}}a_{\rm HB} \approx a_{\rm ej}$,
where $n$ is the $n$'th binary-single interaction, and $N_{\rm ej}$ is the number of interactions it takes to bring the BBH to its ejection value.

For deriving the BBH merger fractions, GW frequencies, and eccentricity distributions, we perform the following calculations at each interaction
step $n$ starting from $n=0$, until the BBH either undergoes a merger or escapes the GC:
We first estimate if the BBH will undergo a 2-body merger, i.e. merge before the next encounter.
For this estimation, we start by calculating the time between successive
binary-single interactions, $t_{\rm bs}$, which can be approximated by $\approx 1/(n_{\rm s}\sigma_{\rm bs}v_{\rm dis})$,
where $n_{\rm s}$ is the number density of single BHs, and $\sigma_{\rm bs}$ is the cross section for a binary-single interaction at step $n$ \citep[\textit{e.g.},][]{2018ApJ...853..140S}.
We then derive the GW-inspiral lifetime of the BBH assuming its eccentricity is $=0$, denoted by $t_{\rm c}$, using the prescriptions from \cite{Peters:1964bc}. From these two derived
time scales, we can then calculate what the minimum eccentricity of the BBH must be for it to undergo a GW merger before its next encounter, denoted by $e_{\rm 2b}$,
which is the solution to the following relation $t_{\rm bs} = t_{\rm c}(1-e_{\rm 2b}^2)^{7/2}$, assuming $e_{\rm 2b}\gg 0$ \citep{Peters:1964bc}. From this follows,
\begin{equation}
e_{\rm 2b} \approx \sqrt{\left(1-\left( {t_{\rm bs}}/{t_{\rm c}}\right)^{2/7} \right)}.
\end{equation}
To now determine if the BBH will actually undergo a 2-body merger inside the GC at this interaction step $n$, we draw a value for the eccentricity of the BBH, $e$, assuming a
thermal distribution $P(e) = 2e$ \citep{Heggie:1975uy}. If the drawn eccentricity is $\geq e_{\rm 2b}$, the BBH will undergo a 2-body merger, and we record its orbital elements.
If the BBH does not merge, i.e. if the drawn eccentricity is $< e_{\rm 2b}$, we then move on to estimate if the BBH instead undergoes a merger during
its next binary-single interaction.

For estimating the probability of a 3-body merger we use the framework first presented in \cite{2014ApJ...784...71S}, in which the binary-single interaction is pictured as
a series of states composed of a binary, referred to as an intermediate state (IMS) binary, and a bound single. As described in \cite{2017arXiv171107452S}, on average about
$N_{\rm IMS} \approx 20$ IMS binaries will form per binary-single interaction, each with a SMA that is about the initial SMA of the target binary
and an eccentricity that is drawn from the thermal distribution $P(e) = 2e$. The probability for a 3-body merger to form during the three-body interaction is
equal to the probability for an IMS binary to undergo a GW merger within the orbital time of the bound single.
To calculate this probability, we first estimate the characteristic pericentre distance an IMS binary must have for it to undergo a GW capture merger during the interaction,
a distance we denote by $r_{\rm 3b}$.
Although this distance changes between each IMS in the three-body interaction \citep{2014ApJ...784...71S},
one finds that on average $r_{\rm 3b}$ is about the distance for which the energy loss through GW emission integrated over one IMS binary orbit
is similar to the initial total energy of the three-body system. From this it follows that,
$r_{\rm 3b} \approx {\mathscr{R}_{\rm m}} \times ({a}/{{\mathscr{R}_{\rm m}}})^{2/7}$, where ${\mathscr{R}_{\rm m}}$ here
denotes the Schwarzschild radius of a BH with mass $m$, and $a$ is the SMA of the target binary, which in our step wise hardening
series equals $\delta^{n}a_{\rm HB}$ for step $n$ \citep[see][]{2017arXiv171107452S}. The minimum eccentricity of an IMS BBH
needed to undergo a GW capture merger during the interaction is then given by,
\begin{equation}
e_{\rm 3b} \approx 1 - r_{\rm 3b}/a.
\end{equation}
As for the 2-body mergers, we then draw a value for the IMS BBH eccentricity from the thermal distribution $2e$. We do this up to $N_{\rm IMS} = 20$ times for each interaction.
If one of the drawn eccentricities is $\geq e_{\rm 3b}$ a 3-body merger has formed and we save its orbital elements.

If neither a 2-body nor a 3-body merger has formed at the considered step $n$, we move on to the next SMA step in the hardening series,
which after the last binary-single interaction is now $\delta^{n+1}a_{\rm HB}$, and redo the above calculations. If no merger has
formed when the BBH SMA falls below $a_{\rm ej}$, we assume the BBH escapes the cluster with a SMA $ = a_{\rm ej}$.
For this BBH we then calculate what its minimum eccentricity must be
for it to merge within in a Hubble time, $e_{\rm Ht}$. We again draw from a thermal distribution in eccentricity, and if the value is $ > e_{\rm Ht}$ we label the
BBH as an ejected BBH merger. 

For this paper we follow $10^6$ such BBHs starting at their $a_{\rm HB}$ from which we then derive BBH
merger fractions, frequency and eccentricity distributions, from the above procedure by going through each of the hardening steps. This allows us to
investigate the role of GR effects in what effectively corresponds to $>10^7$ 2.5 PN binary-single scatterings in just a few seconds. Our results relevant for LISA
are described below.

\section{Eccentric Black Hole Mergers in LISA}\label{sec:Results}

The following results are derived using the method described above, applied to
the scenario for which the interacting BHs are identical, with a mass $30
M_{\odot}$, and for which the population of GCs all have an escape velocity of
$50$ kms$^{-1}$ \citep[\textit{e.g.},][]{1996AJ....112.1487H}. We further
assume that the number density of single BHs, $n_{\rm s}$, in each GC core is $10^{5}$
pc$^{-3}$. This number is highly uncertain; however, one finds that the relative number of 2-body mergers only
scales weakly with density as $n_{\rm s}^{-2/7}$, which follows from \cite{2017arXiv171107452S}. Finally, we note that our chosen
values robustly result in that $\approx 50\%$ of all BBH mergers are in the form of 2-body mergers,
which is in agreement with the recent PN simulations presented in
\cite{2017arXiv171204937R}.

\begin{figure}
\centering
\includegraphics[width=\columnwidth]{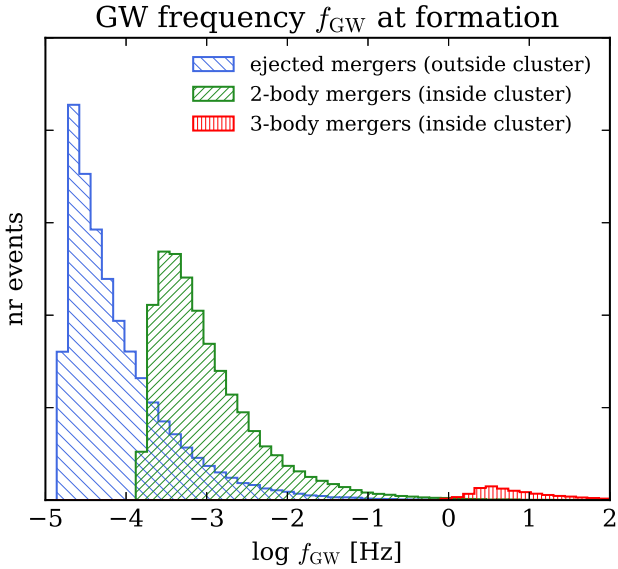}
\caption{Distribution of GW peak frequency $f_{\rm GW}$ at formation for merging BBHs forming dynamically
in GCs. The distributions here are derived using our simple BBH hardening model described in Section \ref{sec:Model}, which combines
an MC approach with the analytical framework presented in \citep{2017arXiv171107452S}.
\emph{Blue:} Distribution of BBHs that are ejected from the GC and merge within a Hubble time (ejected mergers).
\emph{Green:} Distribution of BBHs that merge inside the GC \emph{between} their hardening binary-single interactions (2-body mergers).
\emph{Red:} Distribution of BBHs that merge inside the GC \emph{during} their hardening binary-single interactions (3-body mergers).
The relative contribution from each population depends on the masses of the interacting BHs, the density of single BHs in the GC core, and the escape velocity
of the GC; however, all reasonable values lead to about half of all merging BBHs merging inside the cluster (green/red), where about $5\%$ of merging BBHs form in 3-body mergers.
We emphasize that the 2-body and 3-body merger populations only can be resolved with GR included in the $N$-body modeling.}
\label{fig:fGW0}
\end{figure}

The distributions of peak GW frequency, $f_{\rm GW}$, at the time of formation
of the BBHs that are merging through the three different pathways considered
in this work (ejected merger, 2-body merger, 3-body merger) are shown in
Figure \ref{fig:fGW0}. For this we used the approximation $f_{\rm GW} =
\pi^{-1} \sqrt{{2Gm}/{r_{\rm p}^3}}$, where $r_{\rm p}$ is the pericentre
distance at the time of formation of the merging BBH
\citep[\textit{e.g.},][]{Wen:2003bu, 2017arXiv171107452S}. The ejected mergers (blue)
initially distribute at relatively low $f_{\rm GW}$ with a peak between
$10^{-5}-10^{-4}$ Hz, and will therefore drift through both LISA and LIGO. The
possibility for joint observations have been suggested for such a
population \citep[\textit{e.g.},][]{2016PhRvL.116w1102S, Seto:2016}. The 3-body mergers
(red) all have a much higher initial $f_{\rm GW}$ with a peak between
$10^{0}-10^{1}$ Hz, and will therefore elude the LISA band and form directly
in the proposed DECIGO \citep{2011CQGra..28i4011K, 2018arXiv180206977I}/Tian
Qin \citep{TianQin} band before entering the LIGO band
\citep[\textit{e.g.},][]{2017ApJ...842L...2C, 2018arXiv180208654S}. We note here that
these two distributions are in full agreement with those found in
\cite{2018arXiv180208654S}, in which the distributions were resolved using
full numerical 2.5 PN scatterings using data from the \texttt{MOCCA} code
\citep{Giersz2013, 2017MNRAS.464L..36A}. This validates at least this part of
our framework. The 2-body mergers (green) interestingly distribute between the
ejected and the 3-body mergers, with a peak only slightly below the maximum sensitivity 
region of LISA. 
A proper understanding and modeling of this
population is required for using LISA to determine the origin of BBH mergers.
As stated in Section \ref{sec:BBH mergers in the LISA band}, we note that 
this population has not been studied in this context before.
In Paper II of this series we investigate in detail the GW signatures of these three dynamically formed populations in the LISA band.

\begin{figure}
\centering
\includegraphics[width=\columnwidth]{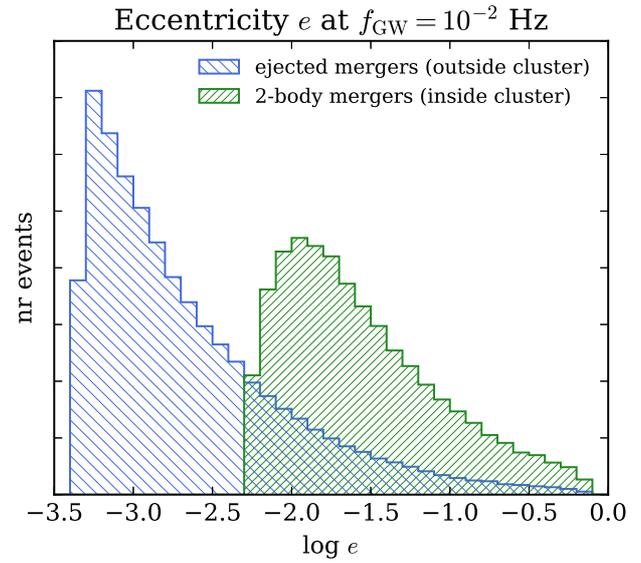}
\caption{Distribution of BBH orbital eccentricity $e$ at $10^{-2}$ Hz derived using all the BBH mergers from the set presented in Figure \ref{fig:fGW0} that
have an initial $f_{\rm GW} < 10^{-2}$ Hz. As the 3-body mergers peak at much higher frequencies, the
considered set is completely dominated by the ejected (blue) and 2-body (green) mergers. As seen, the 2-body mergers dominate the
fraction that will have a resolvable eccentricity ($>0.01$) in the LISA band ($10^{-2}$ Hz). This population will therefore play a key role in
determining the origin of BBH mergers using a LISA-like instrument, as, \textit{e.g.}, field BBHs are expected to be circular to a much higher degree in LISA.}
\label{fig:eccHIST}
\end{figure}

\begin{figure}
\centering
\includegraphics[width=\columnwidth]{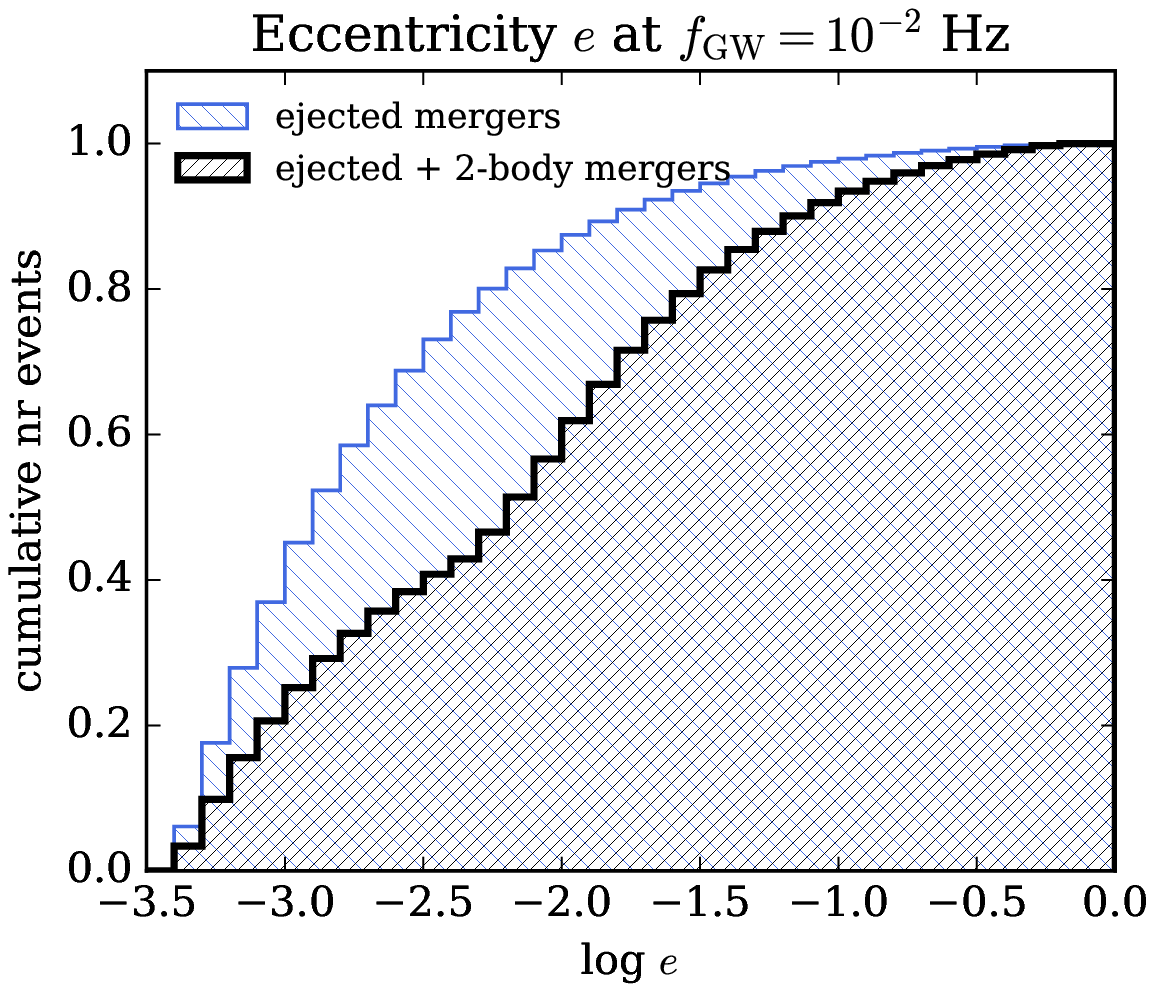}
\caption{Cumulative distribution of the eccentricity distributions shown in Figure \ref{fig:eccHIST}.
When the 2-body mergers are included (black), $\approx 40\%$ of merging BBHs will have an eccentricity $>0.01$ at $10^{-2}$ Hz, near the peak sensitivity of the LISA band. We note that this fraction is about $4$ times higher than recently reported by \citep{2016ApJ...830L..18B}, who effectively only considered the ejected population.
A substantial fraction of eccentric BBH mergers are therefore expected in LISA if the dynamical GC channel contributes to the BBH merger rate. This finding should
be taken into account when optimizing science cases and instrumental designs.}
\label{fig:CUMUeccHIST}
\end{figure}

The eccentricity distribution of the BBH mergers evaluated at $10^{-2}$ Hz,
near the peak of the LISA sensitivity, is shown in Figure \ref{fig:eccHIST}.
To derive this distribution, we use the evolution equations from
\cite{Peters:1964bc} to propagate the initial BBH eccentricity distribution,
with initial peak GW frequency $f_{\rm GW} < 10^{-2}$ Hz, to the value at
$f_{\rm GW} = 10^{-2}$ Hz. As seen, the 2-body mergers, i.e. the
BBHs that merge between encounters inside the GC, completely dominate the
fraction of mergers that will have an eccentricity resolvable by LISA ($e>
0.01$). To clarify this statement, Figure \ref{fig:CUMUeccHIST} shows the
corresponding cumulative distribution. As seen, if only the ejected mergers
are considered (as was effectively done in \cite{2016ApJ...830L..18B}), then
only $\approx 10\%$ will have an eccentricity $> 0.01$ at $10^{-2}$ Hz (blue);
however, when the 2-body mergers are included $\approx 40\%$ of all the
mergers will have an eccentricity $ > 0.01$ (black). This is an important
correction, as some recent studies have argued that eccentric populations
would hint for BBH mergers forming near massive BHs
\citep[\textit{e.g.},][]{2017MNRAS.465.4375N}. Our results show that GCs can
produce eccentric mergers in LISA as well, greatly motivating further and
more detailed studies on systems.

From this we conclude that BBH mergers forming in GCs are expected to lead to
a notable distribution of eccentric sources ($>0.01$) in the LISA band
($10^{-2}$ Hz), with a relative fraction that is significantly higher than
recently reported by \cite{2016ApJ...830L..18B}. This not only shows the
importance of a proper inclusion of GR terms in current $N$-body studies, but
also the bright prospects of observationally distinguishing where and how BBH
mergers form in our Universe with LISA.

\section{Conclusions}\label{sec:Conclusions}

In Paper I of this series, we have explored the role of GR effects in the dynamical evolution of BBHs inside GCs, and found that
the population that merges through GW emission between their hardening binary-single interactions, referred to as 2-body mergers,
all appear with a notable eccentricity ($>0.01$) in the LISA band ($10^{-2}$ Hz). Using a simple MC approach
together with the analytical framework presented in \cite{2017arXiv171107452S},
we find that with the inclusion of these 2-body mergers, $\approx 40\%$ of all BBH mergers from GCs will be eccentric in LISA, which is
$\approx 4$ times more than recently stated by \cite{2016ApJ...830L..18B}, in which only Newtonian gravity was included.

That GCs are expected to have much richer distributions in eccentricity across the LISA band than previously thought,
has important implications for how to observationally distinguish BBH merger channels from each other using LISA, as well
as LIGO \citep[\textit{e.g.},][]{2016PhRvD..94f4020N, 2016ApJ...830L..18B, 2017MNRAS.465.4375N, 2018arXiv180208654S}.
The reason is that different channels will have different eccentricity distributions, e.g. isolated field binaries are believed to have almost circularized
once entering LISA, whereas BBH mergers assembled near massive black holes have been shown to have a notable eccentricity in LISA \citep[\textit{e.g.},][]{2017MNRAS.465.4375N}.

Our results likewise indicate that the background of unresolved sources
observable by LISA, is likely to have a significant fraction of eccentric
sources. Including such a population will lead to changes in the expected
background spectrum, which often is assumed to be dominated by circular BBHs
partly due to the Newtonian results derived in \cite{2016PhRvD..93h4029R},
that we argue greatly underestimates the true fraction of eccentric sources.
In Paper II of this series, we explore the tracks of individually resolvable,
eccentric BBHs through the LISA band as well as the effect of unresolvable
eccentric systems on the gravitational wave background detectable by LISA,
each a result of the GR effects discussed in this paper.

\section*{Acknowledgements}
It is a pleasure to thank M. Giersz, A. Askar, E. Kovetz, and M. Kamionkowski for helpful discussions.
J.S. acknowledges support from the Lyman Spitzer Fellowship.
D.J.D. acknowledges financial
support from NASA through Einstein Postdoctoral Fellowship award
number PF6-170151.
D.J.D. also thanks Adrian Price-Whelan and Lauren Glattly for their hospitality during the conception of this work.



\begin{thebibliography}{}
\makeatletter
\relax
\def\mn@urlcharsother{\let\do\@makeother \do\$\do\&\do\#\do\^\do\_\do\%\do\~}
\def\mn@doi{\begingroup\mn@urlcharsother \@ifnextchar [ {\mn@doi@}
  {\mn@doi@[]}}
\def\mn@doi@[#1]#2{\def\@tempa{#1}\ifx\@tempa\@empty \href
  {http://dx.doi.org/#2} {doi:#2}\else \href {http://dx.doi.org/#2} {#1}\fi
  \endgroup}
\def\mn@eprint#1#2{\mn@eprint@#1:#2::\@nil}
\def\mn@eprint@arXiv#1{\href {http://arxiv.org/abs/#1} {{\tt arXiv:#1}}}
\def\mn@eprint@dblp#1{\href {http://dblp.uni-trier.de/rec/bibtex/#1.xml}
  {dblp:#1}}
\def\mn@eprint@#1:#2:#3:#4\@nil{\def\@tempa {#1}\def\@tempb {#2}\def\@tempc
  {#3}\ifx \@tempc \@empty \let \@tempc \@tempb \let \@tempb \@tempa \fi \ifx
  \@tempb \@empty \def\@tempb {arXiv}\fi \@ifundefined
  {mn@eprint@\@tempb}{\@tempb:\@tempc}{\expandafter \expandafter \csname
  mn@eprint@\@tempb\endcsname \expandafter{\@tempc}}}

\bibitem[\protect\citeauthoryear{{Aarseth} \& {Heggie}}{{Aarseth} \&
  {Heggie}}{1976}]{1976A&A....53..259A}
{Aarseth} S.~J.,  {Heggie} D.~C.,  1976, \aap, \href
  {http://adsabs.harvard.edu/abs/1976A%26A....53..259A} {53, 259}

\bibitem[\protect\citeauthoryear{{Abbott} et~al.,}{{Abbott}
  et~al.}{2016a}]{2016PhRvX...6d1015A}
{Abbott} B.~P.,  et~al., 2016a, \mn@doi [Physical Review X]
  {10.1103/PhysRevX.6.041015}, \href
  {http://adsabs.harvard.edu/abs/2016PhRvX...6d1015A} {6, 041015}

\bibitem[\protect\citeauthoryear{{Abbott} et~al.,}{{Abbott}
  et~al.}{2016b}]{2016PhRvL.116f1102A}
{Abbott} B.~P.,  et~al., 2016b, \mn@doi [Physical Review Letters]
  {10.1103/PhysRevLett.116.061102}, \href
  {http://adsabs.harvard.edu/abs/2016PhRvL.116f1102A} {116, 061102}

\bibitem[\protect\citeauthoryear{{Abbott} et~al.,}{{Abbott}
  et~al.}{2016c}]{2016PhRvL.116x1103A}
{Abbott} B.~P.,  et~al., 2016c, \mn@doi [Physical Review Letters]
  {10.1103/PhysRevLett.116.241103}, \href
  {http://adsabs.harvard.edu/abs/2016PhRvL.116x1103A} {116, 241103}

\bibitem[\protect\citeauthoryear{{Abbott} et~al.,}{{Abbott}
  et~al.}{2017a}]{2017PhRvL.118v1101A}
{Abbott} B.~P.,  et~al., 2017a, \mn@doi [Physical Review Letters]
  {10.1103/PhysRevLett.118.221101}, \href
  {http://adsabs.harvard.edu/abs/2017PhRvL.118v1101A} {118, 221101}

\bibitem[\protect\citeauthoryear{{Abbott} et~al.,}{{Abbott}
  et~al.}{2017b}]{2017PhRvL.119n1101A}
{Abbott} B.~P.,  et~al., 2017b, \mn@doi [Physical Review Letters]
  {10.1103/PhysRevLett.119.141101}, \href
  {http://adsabs.harvard.edu/abs/2017PhRvL.119n1101A} {119, 141101}

\bibitem[\protect\citeauthoryear{{Amaro-Seoane} et~al.,}{{Amaro-Seoane}
  et~al.}{2017}]{2017arXiv170200786A}
{Amaro-Seoane} P.,  et~al., 2017, preprint, \href
  {http://adsabs.harvard.edu/abs/2017arXiv170200786A} {} (\mn@eprint {arXiv}
  {1702.00786})

\bibitem[\protect\citeauthoryear{{Antonini} \& {Rasio}}{{Antonini} \&
  {Rasio}}{2016}]{2016ApJ...831..187A}
{Antonini} F.,  {Rasio} F.~A.,  2016, \mn@doi [\apj]
  {10.3847/0004-637X/831/2/187}, \href
  {http://adsabs.harvard.edu/abs/2016ApJ...831..187A} {831, 187}

\bibitem[\protect\citeauthoryear{{Antonini}, {Rodriguez}, {Petrovich}  \&
  {Fischer}}{{Antonini} et~al.}{2017}]{2017arXiv171107142A}
{Antonini} F.,  {Rodriguez} C.~L.,  {Petrovich} C.,   {Fischer} C.~L.,  2017,
  preprint, \href {http://adsabs.harvard.edu/abs/2017arXiv171107142A} {}
  (\mn@eprint {arXiv} {1711.07142})

\bibitem[\protect\citeauthoryear{{Askar}, {Szkudlarek}, {Gondek-Rosi{\'n}ska},
  {Giersz}  \& {Bulik}}{{Askar} et~al.}{2017}]{2017MNRAS.464L..36A}
{Askar} A.,  {Szkudlarek} M.,  {Gondek-Rosi{\'n}ska} D.,  {Giersz} M.,
  {Bulik} T.,  2017, \mn@doi [\mnras] {10.1093/mnrasl/slw177}, \href
  {http://adsabs.harvard.edu/abs/2017MNRAS.464L..36A} {464, L36}

\bibitem[\protect\citeauthoryear{{Bae}, {Kim}  \& {Lee}}{{Bae}
  et~al.}{2014}]{2014MNRAS.440.2714B}
{Bae} Y.-B.,  {Kim} C.,   {Lee} H.~M.,  2014, \mn@doi [\mnras]
  {10.1093/mnras/stu381}, \href
  {http://adsabs.harvard.edu/abs/2014MNRAS.440.2714B} {440, 2714}

\bibitem[\protect\citeauthoryear{{Banerjee}, {Baumgardt}  \&
  {Kroupa}}{{Banerjee} et~al.}{2010}]{2010MNRAS.402..371B}
{Banerjee} S.,  {Baumgardt} H.,   {Kroupa} P.,  2010, \mn@doi [\mnras]
  {10.1111/j.1365-2966.2009.15880.x}, \href
  {http://adsabs.harvard.edu/abs/2010MNRAS.402..371B} {402, 371}

\bibitem[\protect\citeauthoryear{{Bartos}, {Kocsis}, {Haiman}  \&
  {M{\'a}rka}}{{Bartos} et~al.}{2017}]{2017ApJ...835..165B}
{Bartos} I.,  {Kocsis} B.,  {Haiman} Z.,   {M{\'a}rka} S.,  2017, \mn@doi
  [\apj] {10.3847/1538-4357/835/2/165}, \href
  {http://adsabs.harvard.edu/abs/2017ApJ...835..165B} {835, 165}

\bibitem[\protect\citeauthoryear{{Belczynski}, {Holz}, {Bulik}  \&
  {O'Shaughnessy}}{{Belczynski} et~al.}{2016a}]{2016Natur.534..512B}
{Belczynski} K.,  {Holz} D.~E.,  {Bulik} T.,   {O'Shaughnessy} R.,  2016a,
  \mn@doi [\nat] {10.1038/nature18322}, \href
  {http://adsabs.harvard.edu/abs/2016Natur.534..512B} {534, 512}

\bibitem[\protect\citeauthoryear{{Belczynski}, {Repetto}, {Holz},
  {O'Shaughnessy}, {Bulik}, {Berti}, {Fryer}  \& {Dominik}}{{Belczynski}
  et~al.}{2016b}]{2016ApJ...819..108B}
{Belczynski} K.,  {Repetto} S.,  {Holz} D.~E.,  {O'Shaughnessy} R.,  {Bulik}
  T.,  {Berti} E.,  {Fryer} C.,   {Dominik} M.,  2016b, \mn@doi [\apj]
  {10.3847/0004-637X/819/2/108}, \href
  {http://adsabs.harvard.edu/abs/2016ApJ...819..108B} {819, 108}

\bibitem[\protect\citeauthoryear{{Bird}, {Cholis}, {Mu{\~n}oz},
  {Ali-Ha{\"i}moud}, {Kamionkowski}, {Kovetz}, {Raccanelli}  \& {Riess}}{{Bird}
  et~al.}{2016}]{2016PhRvL.116t1301B}
{Bird} S.,  {Cholis} I.,  {Mu{\~n}oz} J.~B.,  {Ali-Ha{\"i}moud} Y.,
  {Kamionkowski} M.,  {Kovetz} E.~D.,  {Raccanelli} A.,   {Riess} A.~G.,  2016,
  \mn@doi [Physical Review Letters] {10.1103/PhysRevLett.116.201301}, \href
  {http://adsabs.harvard.edu/abs/2016PhRvL.116t1301B} {116, 201301}

\bibitem[\protect\citeauthoryear{{Blanchet}}{{Blanchet}}{2014}]{2014LRR....17....2B}
{Blanchet} L.,  2014, \mn@doi [Living Reviews in Relativity]
  {10.12942/lrr-2014-2}, \href
  {http://adsabs.harvard.edu/abs/2014LRR....17....2B} {17}

\bibitem[\protect\citeauthoryear{{Breivik}, {Rodriguez}, {Larson}, {Kalogera}
  \& {Rasio}}{{Breivik} et~al.}{2016}]{2016ApJ...830L..18B}
{Breivik} K.,  {Rodriguez} C.~L.,  {Larson} S.~L.,  {Kalogera} V.,   {Rasio}
  F.~A.,  2016, \mn@doi [\apjl] {10.3847/2041-8205/830/1/L18}, \href
  {http://adsabs.harvard.edu/abs/2016ApJ...830L..18B} {830, L18}

\bibitem[\protect\citeauthoryear{{Carr}, {K{\"u}hnel}  \& {Sandstad}}{{Carr}
  et~al.}{2016}]{2016PhRvD..94h3504C}
{Carr} B.,  {K{\"u}hnel} F.,   {Sandstad} M.,  2016, \mn@doi [\prd]
  {10.1103/PhysRevD.94.083504}, \href
  {http://adsabs.harvard.edu/abs/2016PhRvD..94h3504C} {94, 083504}

\bibitem[\protect\citeauthoryear{{Chen} \& {Amaro-Seoane}}{{Chen} \&
  {Amaro-Seoane}}{2017}]{2017ApJ...842L...2C}
{Chen} X.,  {Amaro-Seoane} P.,  2017, \mn@doi [\apjl]
  {10.3847/2041-8213/aa74ce}, \href
  {http://adsabs.harvard.edu/abs/2017ApJ...842L...2C} {842, L2}

\bibitem[\protect\citeauthoryear{{Cholis}, {Kovetz}, {Ali-Ha{\"i}moud}, {Bird},
  {Kamionkowski}, {Mu{\~n}oz}  \& {Raccanelli}}{{Cholis}
  et~al.}{2016}]{2016PhRvD..94h4013C}
{Cholis} I.,  {Kovetz} E.~D.,  {Ali-Ha{\"i}moud} Y.,  {Bird} S.,
  {Kamionkowski} M.,  {Mu{\~n}oz} J.~B.,   {Raccanelli} A.,  2016, \mn@doi
  [\prd] {10.1103/PhysRevD.94.084013}, \href
  {http://adsabs.harvard.edu/abs/2016PhRvD..94h4013C} {94, 084013}

\bibitem[\protect\citeauthoryear{{D'Orazio} \& {Loeb}}{{D'Orazio} \&
  {Loeb}}{2017}]{DOrazioLoeb:2017}
{D'Orazio} D.~J.,  {Loeb} A.,  2017, preprint, \href
  {http://adsabs.harvard.edu/abs/2017arXiv170604211D} {} (\mn@eprint {arXiv}
  {1706.04211})

\bibitem[\protect\citeauthoryear{{Dominik}, {Belczynski}, {Fryer}, {Holz},
  {Berti}, {Bulik}, {Mandel}  \& {O'Shaughnessy}}{{Dominik}
  et~al.}{2012}]{2012ApJ...759...52D}
{Dominik} M.,  {Belczynski} K.,  {Fryer} C.,  {Holz} D.~E.,  {Berti} E.,
  {Bulik} T.,  {Mandel} I.,   {O'Shaughnessy} R.,  2012, \mn@doi [\apj]
  {10.1088/0004-637X/759/1/52}, \href
  {http://adsabs.harvard.edu/abs/2012ApJ...759...52D} {759, 52}

\bibitem[\protect\citeauthoryear{{Dominik}, {Belczynski}, {Fryer}, {Holz},
  {Berti}, {Bulik}, {Mandel}  \& {O'Shaughnessy}}{{Dominik}
  et~al.}{2013}]{2013ApJ...779...72D}
{Dominik} M.,  {Belczynski} K.,  {Fryer} C.,  {Holz} D.~E.,  {Berti} E.,
  {Bulik} T.,  {Mandel} I.,   {O'Shaughnessy} R.,  2013, \mn@doi [\apj]
  {10.1088/0004-637X/779/1/72}, \href
  {http://adsabs.harvard.edu/abs/2013ApJ...779...72D} {779, 72}

\bibitem[\protect\citeauthoryear{{Dominik} et~al.,}{{Dominik}
  et~al.}{2015}]{2015ApJ...806..263D}
{Dominik} M.,  et~al., 2015, \mn@doi [\apj] {10.1088/0004-637X/806/2/263},
  \href {http://adsabs.harvard.edu/abs/2015ApJ...806..263D} {806, 263}

\bibitem[\protect\citeauthoryear{{Giersz}, {Heggie}, {Hurley}  \&
  {Hypki}}{{Giersz} et~al.}{2013}]{Giersz2013}
{Giersz} M.,  {Heggie} D.~C.,  {Hurley} J.~R.,   {Hypki} A.,  2013, \mn@doi
  [\mnras] {10.1093/mnras/stt307}, \href
  {http://adsabs.harvard.edu/abs/2013MNRAS.431.2184G} {431, 2184}

\bibitem[\protect\citeauthoryear{G{\"u}ltekin, Miller  \&
  Hamilton}{G{\"u}ltekin et~al.}{2006}]{2006ApJ...640..156G}
G{\"u}ltekin K.,  Miller M.~C.,   Hamilton D.~P.,  2006, \apj, 640, 156

\bibitem[\protect\citeauthoryear{{Harris}}{{Harris}}{1996}]{1996AJ....112.1487H}
{Harris} W.~E.,  1996, \mn@doi [\aj] {10.1086/118116}, \href
  {http://adsabs.harvard.edu/abs/1996AJ....112.1487H} {112, 1487}

\bibitem[\protect\citeauthoryear{Heggie}{Heggie}{1975}]{Heggie:1975uy}
Heggie D.~C.,  1975, \mnras, 173, 729

\bibitem[\protect\citeauthoryear{{Hoang}, {Naoz}, {Kocsis}, {Rasio}  \&
  {Dosopoulou}}{{Hoang} et~al.}{2017}]{2017arXiv170609896H}
{Hoang} B.-M.,  {Naoz} S.,  {Kocsis} B.,  {Rasio} F.~A.,   {Dosopoulou} F.,
  2017, preprint, \href {http://adsabs.harvard.edu/abs/2017arXiv170609896H} {}
  (\mn@eprint {arXiv} {1706.09896})

\bibitem[\protect\citeauthoryear{{Hong} \& {Lee}}{{Hong} \&
  {Lee}}{2015}]{2015MNRAS.448..754H}
{Hong} J.,  {Lee} H.~M.,  2015, \mn@doi [\mnras] {10.1093/mnras/stv035}, \href
  {http://adsabs.harvard.edu/abs/2015MNRAS.448..754H} {448, 754}

\bibitem[\protect\citeauthoryear{Hut \& Bahcall}{Hut \&
  Bahcall}{1983}]{Hut:1983js}
Hut P.,  Bahcall J.~N.,  1983, \apj, 268, 319

\bibitem[\protect\citeauthoryear{{Isoyama}, {Nakano}  \& {Nakamura}}{{Isoyama}
  et~al.}{2018}]{2018arXiv180206977I}
{Isoyama} S.,  {Nakano} H.,   {Nakamura} T.,  2018, preprint, \href
  {http://adsabs.harvard.edu/abs/2018arXiv180206977I} {} (\mn@eprint {arXiv}
  {1802.06977})

\bibitem[\protect\citeauthoryear{{Janiuk}, {Bejger}, {Charzy{\'n}ski}  \&
  {Sukova}}{{Janiuk} et~al.}{2017}]{Janiuk+2017}
{Janiuk} A.,  {Bejger} M.,  {Charzy{\'n}ski} S.,   {Sukova} P.,  2017,
  preprint, \href {http://adsabs.harvard.edu/abs/2017NewA...51....7J} {51, 7}
  (\mn@eprint {arXiv} {1604.07132})

\bibitem[\protect\citeauthoryear{{Kawamura} et~al.,}{{Kawamura}
  et~al.}{2011}]{2011CQGra..28i4011K}
{Kawamura} S.,  et~al., 2011, \mn@doi [Classical and Quantum Gravity]
  {10.1088/0264-9381/28/9/094011}, \href
  {http://adsabs.harvard.edu/abs/2011CQGra..28i4011K} {28, 094011}

\bibitem[\protect\citeauthoryear{{Liu} \& {Lai}}{{Liu} \&
  {Lai}}{2017}]{2017ApJ...846L..11L}
{Liu} B.,  {Lai} D.,  2017, \mn@doi [\apjl] {10.3847/2041-8213/aa8727}, \href
  {http://adsabs.harvard.edu/abs/2017ApJ...846L..11L} {846, L11}

\bibitem[\protect\citeauthoryear{{Loeb}}{{Loeb}}{2016}]{Loeb:2016}
{Loeb} A.,  2016, \mn@doi [\apjl] {10.3847/2041-8205/819/2/L21}, \href
  {http://adsabs.harvard.edu/abs/2016ApJ...819L..21L} {819, L21}

\bibitem[\protect\citeauthoryear{{Luo} et~al.,}{{Luo} et~al.}{2016}]{TianQin}
{Luo} J.,  et~al., 2016, \mn@doi [Classical and Quantum Gravity]
  {10.1088/0264-9381/33/3/035010}, \href
  {http://adsabs.harvard.edu/abs/2016CQGra..33c5010L} {33, 035010}

\bibitem[\protect\citeauthoryear{{McKernan} et~al.,}{{McKernan}
  et~al.}{2017}]{2017arXiv170207818M}
{McKernan} B.,  et~al., 2017, preprint, \href
  {http://adsabs.harvard.edu/abs/2017arXiv170207818M} {} (\mn@eprint {arXiv}
  {1702.07818})

\bibitem[\protect\citeauthoryear{{Nishizawa}, {Berti}, {Klein}  \&
  {Sesana}}{{Nishizawa} et~al.}{2016}]{2016PhRvD..94f4020N}
{Nishizawa} A.,  {Berti} E.,  {Klein} A.,   {Sesana} A.,  2016, \mn@doi [\prd]
  {10.1103/PhysRevD.94.064020}, \href
  {http://adsabs.harvard.edu/abs/2016PhRvD..94f4020N} {94, 064020}

\bibitem[\protect\citeauthoryear{{Nishizawa}, {Sesana}, {Berti}  \&
  {Klein}}{{Nishizawa} et~al.}{2017}]{2017MNRAS.465.4375N}
{Nishizawa} A.,  {Sesana} A.,  {Berti} E.,   {Klein} A.,  2017, \mn@doi
  [\mnras] {10.1093/mnras/stw2993}, \href
  {http://adsabs.harvard.edu/abs/2017MNRAS.465.4375N} {465, 4375}

\bibitem[\protect\citeauthoryear{{O'Leary}, {Kocsis}  \& {Loeb}}{{O'Leary}
  et~al.}{2009}]{2009MNRAS.395.2127O}
{O'Leary} R.~M.,  {Kocsis} B.,   {Loeb} A.,  2009, \mn@doi [\mnras]
  {10.1111/j.1365-2966.2009.14653.x}, \href
  {http://adsabs.harvard.edu/abs/2009MNRAS.395.2127O} {395, 2127}

\bibitem[\protect\citeauthoryear{{Park}, {Kim}, {Lee}, {Bae}  \&
  {Belczynski}}{{Park} et~al.}{2017}]{2017MNRAS.469.4665P}
{Park} D.,  {Kim} C.,  {Lee} H.~M.,  {Bae} Y.-B.,   {Belczynski} K.,  2017,
  \mn@doi [\mnras] {10.1093/mnras/stx1015}, \href
  {http://adsabs.harvard.edu/abs/2017MNRAS.469.4665P} {469, 4665}

\bibitem[\protect\citeauthoryear{Peters}{Peters}{1964}]{Peters:1964bc}
Peters P.,  1964, Phys. Rev., 136, B1224

\bibitem[\protect\citeauthoryear{{Portegies Zwart} \& {McMillan}}{{Portegies
  Zwart} \& {McMillan}}{2000}]{2000ApJ...528L..17P}
{Portegies Zwart} S.~F.,  {McMillan} S.~L.~W.,  2000, \mn@doi [\apjl]
  {10.1086/312422}, \href {http://adsabs.harvard.edu/abs/2000ApJ...528L..17P}
  {528, L17}

\bibitem[\protect\citeauthoryear{{Rodriguez}, {Morscher}, {Pattabiraman},
  {Chatterjee}, {Haster}  \& {Rasio}}{{Rodriguez}
  et~al.}{2015}]{2015PhRvL.115e1101R}
{Rodriguez} C.~L.,  {Morscher} M.,  {Pattabiraman} B.,  {Chatterjee} S.,
  {Haster} C.-J.,   {Rasio} F.~A.,  2015, \mn@doi [Physical Review Letters]
  {10.1103/PhysRevLett.115.051101}, \href
  {http://adsabs.harvard.edu/abs/2015PhRvL.115e1101R} {115, 051101}

\bibitem[\protect\citeauthoryear{{Rodriguez}, {Chatterjee}  \&
  {Rasio}}{{Rodriguez} et~al.}{2016a}]{2016PhRvD..93h4029R}
{Rodriguez} C.~L.,  {Chatterjee} S.,   {Rasio} F.~A.,  2016a, \mn@doi [\prd]
  {10.1103/PhysRevD.93.084029}, \href
  {http://adsabs.harvard.edu/abs/2016PhRvD..93h4029R} {93, 084029}

\bibitem[\protect\citeauthoryear{{Rodriguez}, {Haster}, {Chatterjee},
  {Kalogera}  \& {Rasio}}{{Rodriguez} et~al.}{2016b}]{2016ApJ...824L...8R}
{Rodriguez} C.~L.,  {Haster} C.-J.,  {Chatterjee} S.,  {Kalogera} V.,   {Rasio}
  F.~A.,  2016b, \mn@doi [\apjl] {10.3847/2041-8205/824/1/L8}, \href
  {http://adsabs.harvard.edu/abs/2016ApJ...824L...8R} {824, L8}

\bibitem[\protect\citeauthoryear{{Rodriguez}, {Zevin}, {Pankow}, {Kalogera}  \&
  {Rasio}}{{Rodriguez} et~al.}{2016c}]{2016ApJ...832L...2R}
{Rodriguez} C.~L.,  {Zevin} M.,  {Pankow} C.,  {Kalogera} V.,   {Rasio} F.~A.,
  2016c, \mn@doi [\apjl] {10.3847/2041-8205/832/1/L2}, \href
  {http://adsabs.harvard.edu/abs/2016ApJ...832L...2R} {832, L2}

\bibitem[\protect\citeauthoryear{{Rodriguez}, {Amaro-Seoane}, {Chatterjee}  \&
  {Rasio}}{{Rodriguez} et~al.}{2017}]{2017arXiv171204937R}
{Rodriguez} C.~L.,  {Amaro-Seoane} P.,  {Chatterjee} S.,   {Rasio} F.~A.,
  2017, preprint, \href {http://adsabs.harvard.edu/abs/2017arXiv171204937R} {}
  (\mn@eprint {arXiv} {1712.04937})

\bibitem[\protect\citeauthoryear{{Samsing}}{{Samsing}}{2017}]{2017arXiv171107452S}
{Samsing} J.,  2017, preprint, \href
  {http://adsabs.harvard.edu/abs/2017arXiv171107452S} {} (\mn@eprint {arXiv}
  {1711.07452})

\bibitem[\protect\citeauthoryear{{Samsing}, {MacLeod}  \&
  {Ramirez-Ruiz}}{{Samsing} et~al.}{2014}]{2014ApJ...784...71S}
{Samsing} J.,  {MacLeod} M.,   {Ramirez-Ruiz} E.,  2014, \mn@doi [\apj]
  {10.1088/0004-637X/784/1/71}, \href
  {http://adsabs.harvard.edu/abs/2014ApJ...784...71S} {784, 71}

\bibitem[\protect\citeauthoryear{{Samsing}, {D'Orazio}, {Askar}  \&
  {Giersz}}{{Samsing} et~al.}{2018a}]{2018arXiv180208654S}
{Samsing} J.,  {D'Orazio} D.~J.,  {Askar} A.,   {Giersz} M.,  2018a, preprint,
  \href {http://adsabs.harvard.edu/abs/2018arXiv180208654S} {} (\mn@eprint
  {arXiv} {1802.08654})

\bibitem[\protect\citeauthoryear{{Samsing}, {MacLeod}  \&
  {Ramirez-Ruiz}}{{Samsing} et~al.}{2018b}]{2018ApJ...853..140S}
{Samsing} J.,  {MacLeod} M.,   {Ramirez-Ruiz} E.,  2018b, \mn@doi [\apj]
  {10.3847/1538-4357/aaa715}, \href
  {http://adsabs.harvard.edu/abs/2018ApJ...853..140S} {853, 140}

\bibitem[\protect\citeauthoryear{{Samsing}, {Askar}  \& {Giersz}}{{Samsing}
  et~al.}{2018c}]{2018ApJ...855..124S}
{Samsing} J.,  {Askar} A.,   {Giersz} M.,  2018c, \mn@doi [\apj]
  {10.3847/1538-4357/aaab52}, \href
  {http://adsabs.harvard.edu/abs/2018ApJ...855..124S} {855, 124}

\bibitem[\protect\citeauthoryear{{Sasaki}, {Suyama}, {Tanaka}  \&
  {Yokoyama}}{{Sasaki} et~al.}{2016}]{2016PhRvL.117f1101S}
{Sasaki} M.,  {Suyama} T.,  {Tanaka} T.,   {Yokoyama} S.,  2016, \mn@doi
  [Physical Review Letters] {10.1103/PhysRevLett.117.061101}, \href
  {http://adsabs.harvard.edu/abs/2016PhRvL.117f1101S} {117, 061101}

\bibitem[\protect\citeauthoryear{{Sesana}}{{Sesana}}{2016}]{2016PhRvL.116w1102S}
{Sesana} A.,  2016, \mn@doi [Physical Review Letters]
  {10.1103/PhysRevLett.116.231102}, \href
  {http://adsabs.harvard.edu/abs/2016PhRvL.116w1102S} {116, 231102}

\bibitem[\protect\citeauthoryear{{Seto}}{{Seto}}{2016}]{Seto:2016}
{Seto} N.,  2016, \mn@doi [\mnras] {10.1093/mnrasl/slw060}, \href
  {http://adsabs.harvard.edu/abs/2016MNRAS.460L...1S} {460, L1}

\bibitem[\protect\citeauthoryear{{Stone}, {Metzger}  \& {Haiman}}{{Stone}
  et~al.}{2017}]{2017MNRAS.464..946S}
{Stone} N.~C.,  {Metzger} B.~D.,   {Haiman} Z.,  2017, \mn@doi [\mnras]
  {10.1093/mnras/stw2260}, \href
  {http://adsabs.harvard.edu/abs/2017MNRAS.464..946S} {464, 946}

\bibitem[\protect\citeauthoryear{{Tanikawa}}{{Tanikawa}}{2013}]{2013MNRAS.435.1358T}
{Tanikawa} A.,  2013, \mn@doi [\mnras] {10.1093/mnras/stt1380}, \href
  {http://adsabs.harvard.edu/abs/2013MNRAS.435.1358T} {435, 1358}

\bibitem[\protect\citeauthoryear{{VanLandingham}, {Miller}, {Hamilton}  \&
  {Richardson}}{{VanLandingham} et~al.}{2016}]{2016ApJ...828...77V}
{VanLandingham} J.~H.,  {Miller} M.~C.,  {Hamilton} D.~P.,   {Richardson}
  D.~C.,  2016, \mn@doi [\apj] {10.3847/0004-637X/828/2/77}, \href
  {http://adsabs.harvard.edu/abs/2016ApJ...828...77V} {828, 77}

\bibitem[\protect\citeauthoryear{Wen}{Wen}{2003}]{Wen:2003bu}
Wen L.,  2003, \apj, 598, 419

\bibitem[\protect\citeauthoryear{{Woosley}}{{Woosley}}{2016}]{Woosley:2016}
{Woosley} S.~E.,  2016, \mn@doi [\apjl] {10.3847/2041-8205/824/1/L10}, \href
  {http://adsabs.harvard.edu/abs/2016ApJ...824L..10W} {824, L10}

\makeatother
\end{thebibliography}


\bsp	
\label{lastpage}
\end{document}